\documentclass[aps,pre,preprintnumbers,showpacs,twocolumn,groupedaddress]{revtex4}

\usepackage{amsmath,amssymb} 
\usepackage{graphicx}
\usepackage{psfrag}

\newcommand{\lsbar}{\langle l_s\rangle}

\newcommand{\mbf}{\mathbf}

\begin{document}

\preprint{LMU-ASC 10/07}

\title{Non-affine rubber elasticity for stiff polymer networks} \author{Claus
  Heussinger} \author{Boris Schaefer}\author{Erwin Frey}

\affiliation{Arnold-Sommerfeld Center for Theoretical Physics and
  Center for NanoScience, Department of Physics,
  Ludwig-Maximilians-Universit\"at M\"unchen, Theresienstrasse 37,
  D-80333 M\"unchen, Germany}

\begin{abstract}

  We present a theory for the elasticity of cross-linked stiff polymer
  networks.
  Stiff polymers, unlike their flexible counterparts, are highly anisotropic
  elastic objects. Similar to mechanical beams stiff polymers easily deform in
  bending, while they are much stiffer with respect to tensile forces
  (``stretching'').
  Previous approaches have based network elasticity on the central-force
  stretching mode, in a manner similar to classical rubber elasticity for
  flexible polymers. In contrast, our theory properly accounts for the soft
  bending response inherent to any stiff polymer network.
  A self-consistent effective medium approach is used to calculate the
  macroscopic elastic moduli starting from a microscopic characterization of the
  deformation field in terms of ``floppy modes'' -- low-energy bending
  excitations that retain a high degree of non-affinity.
  The length-scale characterizing the emergent non-affinity is given by the
  ``fiber length'' $l_f$, defined as the scale over which the polymers remain
  straight.
  The calculated scaling properties for the shear modulus are in excellent
  agreement with the results of recent simulations obtained in two-dimensional
  model networks. Furthermore, our theory can be applied to rationalize bulk
  rheological data in reconstituted actin networks.
\end{abstract}

\pacs{62.25.+g, 87.16.Ka, 81.05.Lg} \date{\today}

\maketitle

\section{Introduction}

The elasticity of flexible polymer gels is successfully described by the theory
of rubber elasticity~\cite{rubinstein03}. It ascribes the resistance to
deformation to a reduction of conformational entropy induced by a changing
end-to-end distance of individual polymer strands. In the classic approach,
developed by Kuhn and others~\cite{kuhn}, the magnitude of the deformation of a
single constituent polymer is usually assumed to derive from the macroscopically
induced strain in an affine way. With this assumption the network problem is
reduced to calculating the response of a single chain. In this sense affine
deformations represent a mean-field assumption that neglects spatial
correlations and therefore the coupling between the network structure
(``architecture'') and the mechanical properties of its constituents.

In recent years a different class of cross-linked networks made of
semiflexible or stiff polymers have gained widespread interest. Their
importance for biological systems as the cytoskeleton or
extra-cellular matrix makes understanding their properties highly
rewarding~\cite{bausch06}. Out of the variety of biological stiff
polymers, F-actin has emerged as a model system, which allows precise
{\it in vitro} rheological measurements, for example in determining
the (complex) frequency-dependent shear modulus $G(\omega)$ and in
particular its elastic component, the plateau modulus $G_0$ at
intermediate frequencies. In these experiments various types of
cross-linking proteins are being
used~\cite{wagner06,tha,gar04b,tseng02} and the influence of the
degree of cross-linking on the elastic modulus is investigated.

\begin{figure}[h]
  \begin{center}
    \includegraphics[width=0.9\columnwidth]{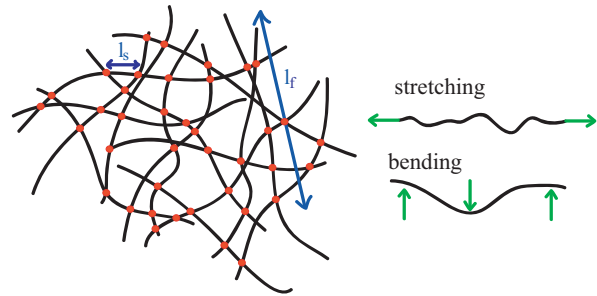}
  \end{center}
  \caption{Sketch of a stiff polymer network with filaments which are straight
    on a scale $l_f$ and where the distances between crosslinks (``polymer
    segments'') on a given filament is denoted by $l_s$.  The response of the
    constituent stiff polymers to external forces is anisotropic with spring
    constants $k_\parallel$ and $k_\perp$, characterizing their resistance to
    stretching and bending deformations, respectively.}
  \label{fig:deformation_modes}
\end{figure}

Stiff polymers, unlike their flexible counterparts, are highly
anisotropic in their elastic response and may be characterized in
terms of two qualitatively different deformation modes (see
Fig.~\ref{fig:deformation_modes})~\cite{kro96,mac95}.  The linear
response to \emph{longitudinal} forces acting parallel to the contour
(stretching/compression), is due to the presence of thermally excited
undulations similar to the (isotropic) stiffness of flexible polymers.
The resulting effective spring constant of a stiff polymer of contour
length $l_s$, $k_\parallel\sim l_p/l_s^4$, depends on the
temperature-dependent persistence length $l_p\sim T^{-1}$, which
indicates the entropic origin.  On the other hand, the resistance of
the polymer to \emph{transverse} forces (bending) is predominantly an
energetic effect, leading to an increase in energy rather than to a
decrease in entropy.  Subsequently, the corresponding spring constant
$k_\perp\sim l_s^{-3}$ is independent of temperature.

The presence of two elementary deformation modes complicates, but also enriches,
the theoretical analysis of stiff polymer networks since it is not obvious which
of the modes, or combination thereof, will dominate the macroscopic elastic
response~\cite{heu06a,heu07architecture}. Leaving aside these difficulties,
recent approaches~\cite{mac95,sto05,shi04} have still adopted straightforward
extensions of rubber elasticity to stiff polymer networks by assuming affine
deformations to be present down to the scale of the individual polymer segment
-- the part of a polymer filament that connects two neighbouring cross-links
(see Fig.\ref{fig:deformation_modes}).  In these models no bending deformations
are present, leaving the stretching mode as the only possible source of
elasticity.

In many systems of interest, however, the large value of the
persistence length $l_p/l \gg 1$ calls this affine approach into
question.  This separation of length scales implies that the bending
mode is in fact soft as compared to the stretching mode, since
$k_\parallel/k_\perp\sim l_p/l \gg 1$.  One would therefore expect the
elastic energy to be dominated by low-energy bending deformations
instead of highly expensive stretching modes~\cite{gib99}.
Interestingly, recent simulations on random fibrous networks have
shown that this is not always the
case~\cite{frey98,wil03,hea03a,heu06a,heu07architecture}.  There, it
was found that in networks with infinitely long filaments (for fixed
density) the soft bending mode is suppressed and the elastic modulus
is entirely given in terms of the stiffer stretching mode, similar to
springs connected in parallel. In contrast, the same simulations
performed in the more realistic situation of finite filament length
have indeed identified an elastic regime that is dominated by soft
bending deformations. The filament length thus strongly influences the
elastic properties and is crucial for understanding the observed
behaviour. The affine theory, by working on the smaller scale of the
polymer segments, is incapable of accounting for these effects.

In the present article we expand on our recent publication~\cite{heu06floppy} to
develop an elastic theory that works on the scale of the whole polymer filament.
The theory naturally explains the presence of a bending dominated regime as well
as its suppression with increasing filament length. It is expected to be
applicable to a broad class of filamentous networks with a soft bending mode.

Similar to the classic theory of rubber elasticity it assumes that the
cross-links adjust to the macroscopically applied strain without showing thermal
fluctuations. In contrast to rubber elasticity, however, the cross-link movemens
are chosen such that the polymer end-to-end distances are kept unchanged. This
automatically avoids energetically highly expensive stretching deformations and
results in elastic moduli that derive from the soft bending mode only.

In the following, we assume that stiff polymers, characterized by
$k_\parallel/k_\perp \gg 1$, effectively behave as if they were strictly
inextensible bars, i.e. having an infinite stretching stiffness $k_\parallel\to
\infty$.  Subsequently, we will construct sets of ``admissible'' cross-link
displacements that respect this inextensibility and thus retain a highly
non-affine character. These displacement modes are referred to as ``floppy
modes''~\cite{thorpe99book}, highlighting the fact that in an equivalent network
of central force springs they would carry no energy. Here, the finite bending
stiffness of the polymers associates an elastic energy to each mode, which we
use to calculate the macroscopic elastic constants of the network.

Section~\ref{sec:floppy-modes} will be concerned with the analysis of networks
in the limit of diverging stretching stiffness $k_\parallel\to\infty$, which
allows us to treat stiff polymers as inextensible bars. We will introduce the
concept of the floppy modes and give an explicit construction valid for a broad
class of network architectures.

In Section~\ref{sec:cost-floppy-modes} we discuss the energy involved with
exciting floppy modes in networks of stiff polymers, characterized by a finite,
but soft, bending stiffness. Specifically, we will develop a theory that allows
to calculate the network elastic constants in a self-consistent manner.

Section~\ref{sec:random-network-2d} is devoted to the specific
architecture of random fibrous networks in a planar geometry (two
dimensonal), where we check our ideas against simulations.

\section{Floppy Modes}\label{sec:floppy-modes}

Here, we are concerned with some general properties of networks of
inextensible bars, so called frameworks. While the bars are assumed to
be perfectly rigid, they are allowed to freely rotate at the
cross-links (``vertices''). In effect, both the stretching and the
bending mode are eliminated, which leaves us with a purely geometric
problem. By applying methods from rigidity theory~\cite{thorpe99book}
we will find that polymer networks when viewed as frameworks are not
rigid and possess zero-energy deformation modes (``floppy modes''),
for which we will give an explicit geometric construction. These
modes, which may be viewed as the analog of the zero-energy shear
modes of regular square lattices, characterize the deformation field
of the network under external strain. By accounting for the finite
bending stiffness of the polymers, they are used to calculate the
elastic energy stored in polymer networks and thus the elastic moduli.

\subsection{Maxwell counting}\label{sec:maxwell-counting}

It has first been realized by Maxwell~\cite{maxwell1864} that a
framework, consisting of $v$ vertices and $b$ bars, can undergo a
transition from a floppy to a rigid state by increasing the
coordination number $z$. Assuming that each bar represents an
independent constraint for the total of $dv$ degrees of freedom in $d$
spatial dimensions, Maxwell derived the condition $b-dv=0$ determining
the rigidity transition. As the number of vertices can be rewritten in
terms of the coordination number as $v=2b/z$, this immediately yields
a critical coordination of $z_c=2d$. According to this simple {\it
  Maxwell counting rule}, frameworks are rigid, whenever their
vertices have more than $z_c$ neighbors, while they will be floppy and
allow for internal rearrangements otherwise.

With regard to stiff polymer networks this transition may be used to
set up a classification where the elastic energy is dominated by
either bending or stretching modes. While for $z<z_c$ bending modes
can stabilize the otherwise floppy (zero-energy) central-force
network, they only provide minor contributions to the energy once
$z>z_c$. The honeycomb lattice in 2d, for example, has a coordination
of $z=3$ and is therefore bending dominated, while the triangular
lattice with $z=6$ is clearly rigid and therefore stretching
dominated. Imposing a deformation necessarily leads to the stretching
of bonds.  A particular case is the square lattice in two dimensions,
which has precisely the critical coordination $z=z_c=4$. Although
being floppy with respect to shear deformations, the network may be
stabilised by introducing suitable boundary constraints or by adding
additional bars along the diagonals of some of the squares. It turns
out that in the limit of infinite system size a non-extensive number
of diagonal bars (which scales as the square root of the system size)
is needed to stabilise the network~\cite{des01}.

Maxwell-counting is only approximate, since one can always add
redundant bars that do not constrain any degrees of freedom. This
effect is taken into account by the modified Maxwell relation $b-dv =
s-m$~\cite{calladine78}. In this picture redundant bars create
overconstrained regions where a total of $s$ states of
self-equilibrated internal stresses may exist. In general, a state of
self-stress is defined as a set of bar tensions that is in static
equilibrium with zero external force applied. At the same time
underconstrained regions arise that allow for $m$ zero energy
deformation modes, i.e. internal rearrangments that can be accomodated
without changing the lengths of any of the bars to first order in the
magnitude of the imposed strain. These are usually referred to as
mechanisms or floppy modes.

In principle, the floppy modes of a pin-jointed structure may be found by
studying the kinematic matrix $\mathbf{C}$ which relates vertex displacements
$\mathbf{d}$ to segment extensions $\mathbf{e=Cd}~$\cite{pellegrino86}. The
kinematic matrix thus constitutes a \emph{linear} relation between displacements
and extensions, which is only true for infinitesimally small displacements. The
entries to the matrix can then easily be identified by considering the extension
of a single bar oriented (in two dimensions) at an angle $\phi$ to the
horizontal. For given displacements $\mathbf{d}_i=(u_i,v_i)$ at the two vertices
$i=1,2$ the extension is found as
\begin{equation}\label{eq:kinematicMatrix}
  e=(u_2-u_1)\cos\phi+(v_2-v_1)\sin\phi\,.
\end{equation}
The floppy modes then correspond to those vertex displacements that do
not lead to any extensions in the bars. This amounts to calculating
the null-space of the matrix, i.e. $\mathbf{Cd_0=0}$.

\begin{figure}[h]
  \begin{center}
    \includegraphics[width=0.9\columnwidth]{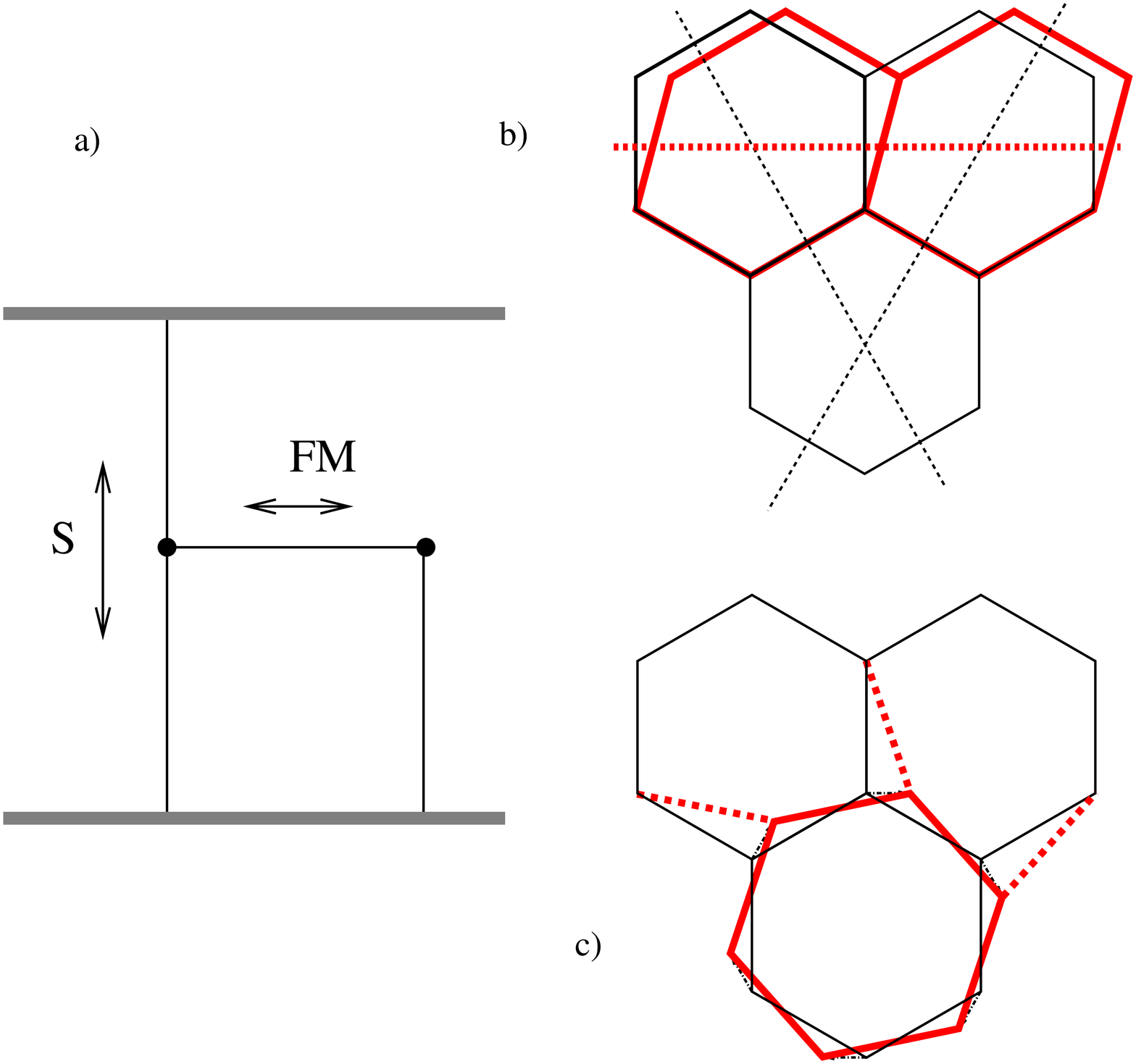}
  \end{center}
  \caption{Illustration of the floppy modes. The ``chair'' in a) has 
    one floppy mode (FM) corresponding to the axial movement of the
    horizontal bar, as well as one state of self-stress (S) located in
    the two vertical bars. The floppy modes of the honeycomb lattice
    may be constructed, b) from the global shear deformations along
    any of the three dashed lines (as well as their parallels), or, c)
    from localized librations.}
  \label{fig:floppymodes}
\end{figure}

An elementary but illustrative example of a bar/joint network (adopted
from~Ref.~\cite{pellegrino93}) is the ``chair'' shown in
Fig.~\ref{fig:floppymodes}a.  Having $b=4$ bars and $v=2$ vertices,
Maxwell's counting rule would imply that the structure is marginally
rigid. Actually, there is also one floppy mode $m=1$ as well as one
state of self-stress $s=1$.  The former corresponds to the
(infinitesimal) movement of the horizontal bar forming the seat, while
the latter corresponds to a tension in the two vertical bars making
the back.

For regular systems it is sometimes possible to guess the modes.
Consider, for the purpose of illustration, a honeycomb lattice in two
dimensions, where a coordination of $z=3$ implies $b-2v=-v/2$. There
is, accordingly, half a floppy mode per vertex. These modes are most
easily identified with shear deformations along lines of symmetry
(Fig.~\ref{fig:floppymodes}b). Probing the shear response of the
honeycomb along a given direction will cause each of the $N$ layers of
cells to be displaced by a small amount $\delta$, which eventually has
to add up to the externally imposed deformation $\Delta = N\delta$.
Thus, there is ``sharing'' of the deformations between the individual
cells and each layer contributes a small amount to fulfilling the
constraints imposed by the macroscopical strain field.  In other
words, the deformation field in the honeycomb lattice is affine down
to the scale of the individual cell, which experiences deformations
$\delta=\Delta/N\propto l_{\rm cell}$ proportional to its own size.

Another possibility to construct the floppy modes of the honeycomb
network is given by the librations of individual
hexagons~\cite{alexander98} (see Fig.~\ref{fig:floppymodes}c). These
librations are, in contrast to the shear displacements, localized
modes that are confined to a single cell and its immediate
surroundings. Since there is one libration per cell and each of the
six corresponding vertices belongs to three cells, this also makes one
mode for every two vertices.

\subsection{Floppy modes of stiff polymer
  networks}\label{sec:floppy-modes-fibrous}

Proteins used to cross-link stiff polymers into networks often have
only two heads~\cite{win05} such that there can only be two-, three-
and four-fold connected vertices. The average coordination number in
stiff polymer networks is therefore $z<4$, which would place it below
the rigidity transition and render the network bending dominated.

In contrast to the very regular structures discussed above, stiff polymer
networks are usually highly random. Nevertheless, as we will see below, the
floppy modes can be constructed quite easily on scales $l_f$ over which the
undeformed polymers can be assumed to be represented by straight fibers. For
isolated polymers the length-scale $l_f$ can be identified with the persistence
length $l_p$, while in networks the origin may be different and for example a
consequence of the network generating process itself. It is the presence of the
length-scale $l_f$ which renders the structure of stiff polymer networks
qualitatively different from flexible polymer gels. The resulting fibrous
appearance may be inspected in the figures of
Refs.~\cite{lieleg,gotter96,collet05pnas}. We have recently argued that the
``fiber length'' $l_f$ plays the role of the size of an effective unit
cell~\cite{heu07architecture}. This has to be contrasted to flexible polymer
gels, where the unit size is set by the mesh-size. In the following we use the
word ``fiber'' in connection with the length $l_f$ over which the polymer
remains straight. In later sections we will introduce a simple model system
where fiber and polymer length are equal.

\begin{figure}[t]
 \begin{center}
   \includegraphics[width=0.9\columnwidth]{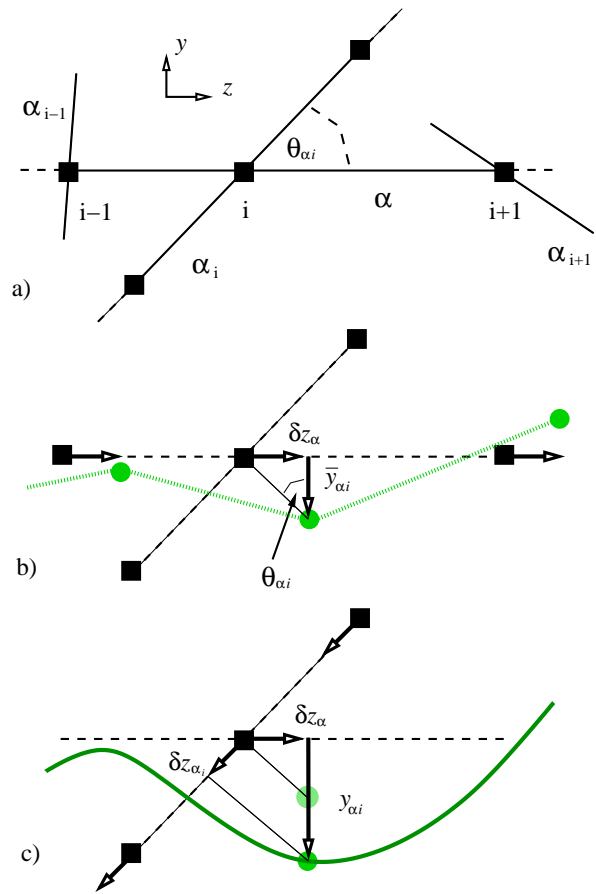}
\end{center}
\caption{Construction of a floppy mode starting from the initial geometry as
  drawn in a). In b), the horizontal fiber is moved, while the surrounding
  fibers remain in their original positions.  This leads to the new cross-link
  positions (green circles) with transverse deflection $\bar y_{\alpha, i} =
  -\cot\theta_{\alpha, i} \, \delta\!z_\alpha$ (Eq.~(\ref{eq:bary})).  The
  component $\mbf{\bar y}_{\alpha, i}^\perp$ (not drawn) is oriented
  perpendicular to the plane of the two fibers. In c), also the secondary fiber
  is moved, such that the cross-link is now deflected according to $y_{\alpha,i}
  = \bar y_{\alpha, i}+\delta\!z_{\alpha_i}/\sin\theta_{\alpha, i}$
  (Eq.~(\ref{eq:bary_total})).  The solid green line represents the actual
  contour of the deformed fiber obtained by minimizing the bending energy along
  the entire fiber (see Eq.~(\ref{eq:w0})).}
  \label{fig:fiberfloppy3d}
\end{figure}

Consider (see Fig.~\ref{fig:fiberfloppy3d}a,b) a single (primary) fiber $\alpha$
of length $l_f$, which may be part of a longer polymer. It is imbedded into a
network of other (secondary, tertiary,~$\ldots$) fibers. Secondary fibers
$\alpha_i$ are assumed to intersect the primary fiber $\alpha$ at the crosslinks
$i=1\ldots n_{\rm cl}$, while tertiary fibers only intersect secondary fibers,
e.t.c.
The floppy-mode construction proceeds in two stages during which only
the cross-links on the primary fiber are being moved.  The rest of the
network, in particular the neighbouring secondary filaments will
remain static such that the floppy mode stays highly localized similar
to the librations of the hexagons discussed above.  In the first step,
we perform a small axial displacement $\delta\!z_\alpha$ of the
primary fiber $\alpha$ as a whole. The axial movement of the
cross-links pertaining to this fiber induces a change in length of all
neighbouring segments on the crossing secondary fibers. In the second
step, therefore, one has to account for the length constraints on
these segments by introducing cross-link deflections
$\mathbf{\bar{y}}_{\alpha, i}$ transverse to the primary fiber. It
turns out that to {\it first order} in $\delta\!z_\alpha$ all segment
lengths can be kept at their repose length by choosing
\begin{equation}\label{eq:bary}
  \mathbf{\bar{y}}_{\alpha, i} = -\delta\!z_\alpha 
  \cot\theta_{\alpha, i} \, \mathbf{\hat e}_{\alpha,i} 
  +\mathbf{\bar y}_{\alpha, i}^{\perp}\,, 
\end{equation}
where $\theta_{\alpha, i}$ is the angle between the two crossing fibers $\alpha$
and $\alpha_i$ at crosslink $i$.  We denote by $\mathbf{\hat e}_{\alpha,i}$ a
unit vector transverse to the primary fiber lying in the plane spanned by the
two fibers, and by $\mathbf{\bar y}_{\alpha, i}^{\perp}$ an arbitrary vector
perpendicular to this plane.

We would like to emphasize that the construction only works for infinitesimal
$\delta\!z_\alpha$, while finite displacements necessarily lead to changes in
bond lengths and therefore to stretching of bonds.  As will be explained in more
detail in Section~\ref{sec:nonlinear} this has dramatic consequences on the
nonlinear elasticity of the network, leading to strong strain stiffening
behaviour.

The construction can be performed for any of the $\alpha = 1, \ldots , N_f$
fibers, such that precisely $N_f$ floppy modes are identified this
way~\footnote{This exhausts all possible floppy modes in two spatial dimensions,
  while in three dimensions additional modes are found that relate to the
  deflection $\mathbf{y}^\perp$ of individual cross-links.}. For the mode
localized around fiber $\alpha$ one may define a vector ${\cal Y}_\alpha =
(0\ldots,\mathbf{\bar y}_{\alpha,1},\ldots, \mathbf{\bar y}_{\alpha, n_{\rm
    cl}},0,\ldots)$, where the deflections of all crosslinks in the network are
combined. It has nonzero components only at crosslinks belonging to fiber
$\alpha$. With respect to the standard vector scalar product one can then show
that the set of floppy modes $\{\cal Y_\alpha\}$ is linearly independent,
however, not orthogonal.  Since a given crosslink $i$ always belongs to two
filaments at the same time, there is obviously a coupling between the two
corresponding modes.

In analogy to the shear modes of the honeycomb lattice one may also construct
extended floppy modes for the fibrous network by superposition of the different
$\cal Y_\alpha$. This amounts to relaxing the constraint that a single fiber
moves in a static environment where neighbouring fibers remain fixed to their
initial positions. Instead, a different $\delta\!z_\alpha$ is assigned to each
of the $\alpha=1, \ldots, N_f$ fibers. In this case, we find (see
Fig.~\ref{fig:fiberfloppy3d}c) that the crosslink deflection $\mathbf{\bar
  y}_{\alpha, i}$ of Eq.~(\ref{eq:bary}) has to be modified by a term
$\delta\!z_{\alpha_i}/\sin\theta_{\alpha, i}$ due to the additional movement
$\delta\!z_{\alpha_i}$ of the neighbouring filament $\alpha_i$ at crosslink
$i$. This amounts to the overall deflection
\begin{equation}\label{eq:bary_total}
  \mathbf{y}_{\alpha,i} = \mathbf{\bar y}_{\alpha,i}
  +\frac{\delta\!z_{\alpha_i}}{\sin\theta_{\alpha, i}}\mathbf{\hat e}_{\alpha, i}\,.
\end{equation}

\begin{figure}[t]
 \begin{center}
   \includegraphics[width=0.8\columnwidth,height=0.8\columnwidth]
   {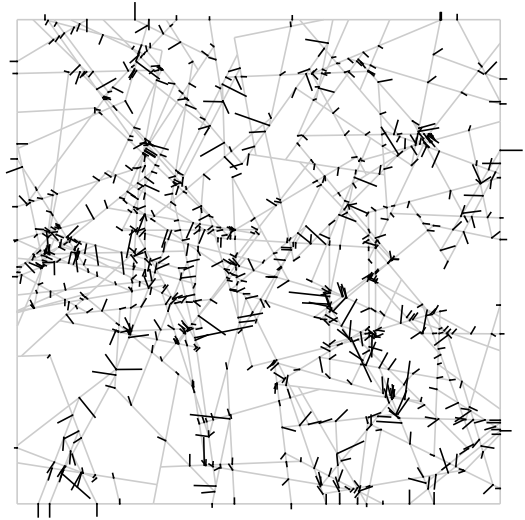}
\end{center}
\caption{Floppy mode of a random fiber network with $225$ fibers (see
  Sect.\ref{sec:random-network-2d}). A fiber has the length of one
  third of the system size. Grey lines represent the network, black
  lines the floppy mode displacements. Note, that the overall ampitude
  of the mode is arbitrary.}
  \label{fig:ortho_fm}
\end{figure}

\begin{figure}[t]
 \begin{center}
   \includegraphics[width=0.8\columnwidth]{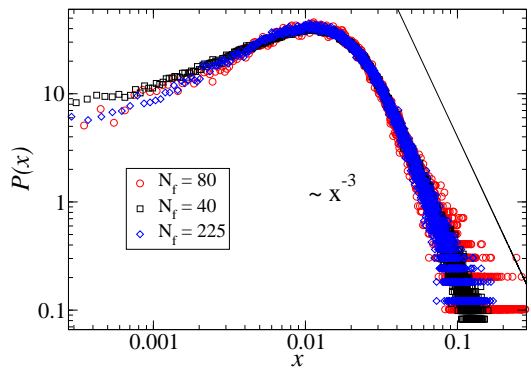}
\end{center}
\caption{Normalized probability distributions of absolute values of 
  floppy mode displacements, as shown in Fig.~\ref{fig:ortho_fm}. The
  distributions for different fiber numbers $N_f$ can be rescaled on a
  single master curve by changing the overall amplitude of the modes.}
  \label{fig:svd_pdf}
\end{figure}

For the particular architecture of a random fiber network in two dimensions
(2d), to be introduced below (see Section~\ref{sec:random-network-2d}), we have
also obtained an orthornormal set of floppy modes. The appropriate values of
$\mathbf{y}_{\alpha,i}$ and $\delta\!z_\alpha$ are found by performing a
singular value decomposition of the compatibility matrix $\mathbf{C}$. One of
the modes is visualized in Fig.~\ref{fig:ortho_fm}, where the black lines
indicate the floppy-mode displacements of the crosslinks. One remarkable
property is the heterogeneous distribution of amplitudes $x$, which leads to
polynomial tails in the probability distribution, $P(x)\sim x^{-3}$ (see
Fig.~\ref{fig:svd_pdf}).  The exponent is a direct consequence of the random
orientation of the filaments which induces a probability distribution of angles
$\theta$ between two intersecting filaments, $P(\theta) \sim
\sin\theta$~\footnote{The probability of intersection of two fibers with
  relative angle $\theta$ is proportional to the excluded volume $A=\sin\theta
  l_f^2/2$.}. By a transformation of variables to the floppy mode deflection
$\bar y\sim \cot\theta$ (and thus to $x$) one finds a distribution $P(\bar
y)\sim \sin^3\theta(\bar y)\to \bar y^{-3}$, where the latter limit corresponds
to large $\bar y\gg 1$.

\section{Energy of floppy modes}\label{sec:cost-floppy-modes}

Having constructed the floppy modes of the fibrous polymer network we now
proceed to determine the energy associated with their excitation.

\subsection{Effective medium theory}

We calculate the response of the network to the axial movement
$\delta\!z_\alpha$ of only one fiber $\alpha$. This proceeds by first
calculating the energy stored in the primary fiber, given that the surrounding
fibers are not allowed to move.  Subsequently, we will relax this constraint and
also account for the energy induced in the secondary (tertiary, $\ldots$)
fibers.  We will find that infinitely many levels of neighboring fibers can be
included by a formal resummation of the energy terms in a Cayley-tree
approximation.

Quite generally, the bending energy of a polymer fiber with weakly undulating
contour can be written as
\begin{equation}\label{eq:fiber_bending}
  W_b[\mathbf{y_\alpha}] = \frac{\kappa}{2}\int_0^{l_f} 
                    \left(\frac{d^2\mathbf{y_\alpha}}{ds^2}\right)^2ds\,,
\end{equation}
where $\mbf{y}_\alpha = \mathbf{y}(s_\alpha)$ denotes the transverse deflection
at point $s_\alpha$ along the backbone of the polymer $\alpha$. The bending
rigidity $\kappa$ is related to the peristence length by $\kappa = l_p k_B T$.
The actual value of the energy contained in a localized floppy mode of amplitude
$\delta\!z_\alpha$ can then be found by minimizing $W_b$ for the given set of
crosslink positions $\mathbf{y}(s_{\alpha, i}) \overset{!}{=} \mathbf{\bar
  y}_{\alpha,i}$ (taken from Eq.~(\ref{eq:bary})), and gives
\begin{equation}\label{eq:w0}
  W_0(\delta\!z_\alpha) = \min_{\mathbf{y}(s_\alpha), 
    \, \mathbf{y}(s_{\alpha, i})
    =\mathbf{\bar y}_{\alpha,i}} W_b[\mathbf{y_\alpha}]\,.
\end{equation} 
Technically, this is achieved by performing a cubic spline interpolation through
the set of points $\{(s_{\alpha, i},\mathbf{\bar y}_{\alpha,
  i})\}_{i=1,\ldots,n_{cl}}$.  This can be shown to be equivalent to the
minimization of the bending energy in
Eq.~(\ref{eq:fiber_bending})~\cite{stoer02}.  As $d^3\mathbf{y}_{\alpha,i}/ds^3$
is proportional to the transverse force in the fiber, the discontinuities of the
cubic spline in its third derivative reflect the external transverse force that
is needed to keep the fiber in its deformed shape.


Now, we assume that neighbouring secondary fibers are free to react to
the movement of the primary fiber. This allows the fiber deflection
$\mbf{y}_{\alpha,i}$ to be different from $\mbf{\bar y}_{\alpha,i}$
and may lead to a reduction of the bending energy on the primary
fiber, however, at the cost of deforming the surrounding, i.e. by
spreading the mode to the neighbouring filaments. The amplitudes of
the so generated secondary floppy modes may be found from solving
Eq.~(\ref{eq:bary_total}) for $\delta\!z_{\alpha_i}$. We thus find
\begin{equation}\label{eq:z_secondary}
  \delta\!z_{\alpha_i} = \sin\theta_{\alpha, i} \, 
  (\mbf{y}_{\alpha, i} - \mbf{\bar y}_{\alpha, i}) \cdot
  \mbf{\hat e}_{\alpha, i}\,.
\end{equation}
which highlights the fact that a secondary mode of amplitude
$\delta\!z_{\alpha_i}$ occurs when the actual transverse deflection
$\mathbf{y}_{\alpha, i}$ is different from the floppy-mode
prescription $\mathbf{\bar y}_{\alpha, i}$. Furthermore, due to the
scalar product with $\mbf{\hat e}_{\alpha, i}$, the displacement
$\mbf{y}_i^\perp$ perpendicular to the plane defined by the two
intersecting polymers does not contribute, $\mbf{y}_{\alpha,
  i}^\perp\cdot \mbf{\hat e}_{\alpha, i} = 0$.

With Eq.~(\ref{eq:z_secondary}) we find that Eq.~(\ref{eq:w0}) has to be
modified by the bending energy contribution $W_0$ from the neighbouring
filaments giving
\begin{equation}\label{eq:w1}
  W_1(\delta\! z_\alpha) 
  = \min_{\mbf{y}(s_\alpha)}\left(W_b[\mbf{y_\alpha}]+\sum_{i=1}^{n_{\rm cl}}
    W_0(\delta\!z_{\alpha_i})\right)\,. 
\end{equation}
Unlike in Eq.~(\ref{eq:w0}), where the crosslink variables on the primary fiber
were constrained to be $\mbf{y}(s_{\alpha, i})=\mbf{\bar y}_{\alpha, i}$, here
they remain unconstrained and move such that the total energy, deriving from
both primary and secondary fibers, is minimized. Note, however, that the
deflections on the secondary fibers are still constrained and given by
Eq.~(\ref{eq:bary}), $\mbf{y}(s_{\alpha_i,j}) = \mbf{\bar y}_{\alpha_i,j}$. This
may be corrected for by taking into account further levels of filaments
(tertiary, $\ldots$), thus defining a sequence of energies
$(W_0,W_1,W_2,...W_\infty)$ the fixed point of which is found by substituting on
both sides of Eq.~(\ref{eq:w1}) one and the same asymptotic function $W_\infty$.

Since the resulting expression still depends on the quenched random
network structure in a complicated way, we have recently proposed an
effective medium approximation that uses the averaged
$\langle[W_\infty]\rangle \equiv W$ instead~\cite{heu06floppy}. For
reasons that will immediately become apparent we have defined two
averaging procedures. The angular brackets $\langle.\rangle$ denote
averaging over the random variables on the primary fiber, the
crosslink positions $s_{\alpha, i}$ and angles $\theta_{\alpha, i}$. The
probability distributions of these variables provide the most
important characterization of the architecture of the network. The
brackets $[.]$ denote averaging with respect to the remaining
randomness in the subsequent hierarchies of fibers. Mathematically,
the effective medium approximation is implemented by interchanging
this latter average with the minimization operation. Physically, this
amounts to assuming that one and the same medium ($W$) is felt by all
the crosslinks on the primary fiber. One thus arrives at the final
equation
\begin{equation}\label{eq:selfconsEn}
  W(\delta\!z_\alpha) =  \left\langle \min_{\mbf{y}(s_\alpha)}
    \left( W_b[y_\alpha] + \sum_{i=1}^{n_{\rm
          cl}} W(\delta\!z_{\alpha_i})\right)\right\rangle     \,,
\end{equation}
where $\delta\!z_{\alpha_i}$ is given by Eq.~(\ref{eq:z_secondary}).
In principle, Eq.~(\ref{eq:selfconsEn}) has to be solved
self-consistently for the function $W(x)$. Since we are concerned with
small displacements only, the energy may be expanded to harmonic order
as $W(x)=kx^2/2$, which gives
\begin{equation}\label{eq:Wharmonic}
  W(\delta\!z_{\alpha_i}) = \frac{1}{2} k 
  \sin^2\theta_{\alpha, i} \left( y_{\alpha, i}-\bar y_{\alpha, i}\right)^2\,,
\end{equation}
where we defined $y_{\alpha, i}=\mbf{y}_{\alpha, i} \cdot \mbf{\hat
  e}_{\alpha, i}$ and similar for $\bar y_{\alpha, i}$.  With this
parametrization Eq.(\ref{eq:selfconsEn}) has to be solved for the
single unknown parameter $k$.

Eq.~(\ref{eq:selfconsEn}) can be interpreted as follows. The total energy stored
in the network upon axially moving a single fiber the amount $\delta\!z_\alpha$
has two contributions. The first term, corresponding to the bending energy of
the primary fiber, $W_b$, dominates if the crosslinks follow the local
floppy-mode associated with $\delta\!z_\alpha$, such that
$\delta\!z_{\alpha_i}\approx 0$ (${y}_{\alpha, i} \approx{\bar y}_{\alpha, i}$).
On the other hand, the energy is mainly stored in the surrounding medium if the
crosslinks deviate strongly from the floppy-mode, ${y}_{\alpha, i} \approx 0$,
in which case the bending energy vanishes, $W_b\approx 0$. Since medium
deformations can only occur in the form of floppy-modes, the stiffness $k$ of
the medium is the same as the stiffness of the fiber.  This allows to solve the
equation self-consistently, which can easily be done numerically as will be
explained in the appendix. There, we will also solve Eq.~(\ref{eq:selfconsEn})
for some exemplary network structures.

\begin{figure}[t]
 \begin{center}
   \includegraphics[width=0.9\columnwidth]{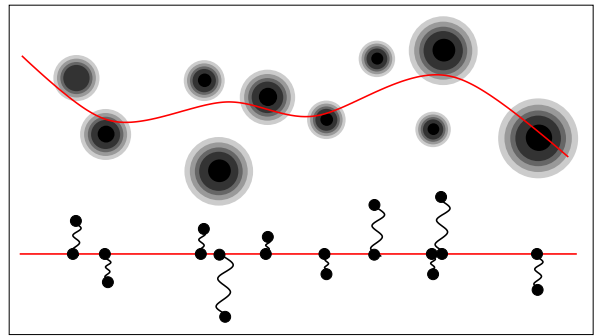}
\end{center}
\caption{The energy in Eq.~(\ref{eq:selfconsEn}) is that of a stiff polymer
  attached to springs of variable stiffness. It may therefore be
  interpreted as a polymer in a random potential. The potential is
  attractive and localized at pinning sites given by
  Eq.~(\ref{eq:bary}).}
  \label{fig:randompotential}
\end{figure}

It is worth mentioning that Eqs.~(\ref{eq:selfconsEn}) and (\ref{eq:Wharmonic})
may be interpreted as the zero temperature limit (or the saddle-point
approximation) to a fluctuating stiff polymer in a random array of harmonic
pinning sites with stiffnesses given by $k\sin^2\theta_{\alpha, i}$ (see
Fig.~\ref{fig:randompotential}).  Compared to the ``bare floppion'' defined by
the Eqs.(\ref{eq:bary})-(\ref{eq:w0}), the excitation given by
Eq.~(\ref{eq:selfconsEn}) is ``dressed'' and incorporates the interactions with
the medium on a Cayley-tree level.

\subsection{Elastic Modulus}\label{sec:deformation-field}

In principle, the elastic modulus can be found by minimizing the energy,
consisting of contributions of the type of Eq.~(\ref{eq:w0}) from each of the
$N_f$ fibers, with respect to the variables $\delta\!z_\alpha$~\footnote{By
  minimizing the energy, instead of calculating the partition function
  associated with the variables $\delta\!z_\alpha$, one neglects fluctuations of
  the crosslinks.}.  Compared to the full problem of having to minimize the
energy with respect to \emph{all} degrees of freedom, that is all $N_{\rm
  cl}\sim N_fn_{\rm cl}$ crosslink coordinates, this is only a minimization
with respect to $N_{f}\ll N_{\rm cl}$ variables. Still, this poses a challenging
quenched disorder problem which can only be tackled numerically.

Here we reduce the calculation to an effective single-fiber problem, by making a
simplifying assumption about the magnitude of the individual $\delta\!z_\alpha$.
We assume, that the fiber centers-of-mass $\mathbf{r}^\alpha_\text{cm}
=(X_\alpha, Y_\alpha, Z_\alpha)$ follow the macroscopic strain field in an
affine way, just as the centers of the hexagons did in the honeycomb lattice.
This is equivalent to assuming that the displacement field is affine on the
scale of the fiber length $l_f$.  Note, however, that this does by no means
imply that the elastic elements themselves undergo affine deformations, as will
become clear below~\footnote{Similarly, the elements of the honeycomb unit cell
  do not undergo affine deformations~\cite{gib99}, even though the unit cell as
  a whole does.}.  For a given macroscopic shear $\gamma \equiv \gamma_{xy}$ we
find $\boldsymbol{\delta r}^\alpha_\text{cm} = \gamma Y_\alpha \mathbf{\hat
  e}_x$ and thus
\begin{equation}\label{eq:affineDeformCM}
\delta\!z_{\alpha} = \gamma Y_\alpha\cos\phi_\alpha\,,
\end{equation}
which is just the projection of the affine displacement on the fiber
axis, oriented at an angle $\phi_\alpha\in[-\pi/2,\pi/2]$ with respect
to the $x$-axis.  Using Eqs.~(\ref{eq:affineDeformCM}) and
(\ref{eq:bary}) one can write Eq.~(\ref{eq:bary_total}) as
\begin{equation}\label{eq:bary_network}
  \mbf{y}_{\alpha, i} = -\delta\!z_\alpha^{\rm rel} 
  \cot\theta_{\alpha, i} \mathbf{\hat e}_{\alpha, i}
  +\mathbf{\bar y}_{\alpha, i}^{\perp}\,,
\end{equation}
where we have defined
\begin{equation}\label{eq:deltaz_network}
  \delta\!z_\alpha^{\rm rel} =
  \gamma
   \left(
    Y_\alpha \cos\phi_\alpha - 
    Y_{\alpha_i}\frac{\cos(\theta_{\alpha, i}+\phi_\alpha)}
                   {\cos\theta_{\alpha,i}}
   \right)\,. 
\end{equation}
Upon comparison of Eq.~(\ref{eq:bary_network}) with Eq.~(\ref{eq:bary}) one may
interpret $\delta\!z_\alpha^{\rm rel}$ as specifying the movement of the primary
fiber \emph{relative} to its surrounding. Note, however, that this relative
displacement $\delta\!z_\alpha^{\rm rel}$ depends on the orientations
$\phi_\alpha$ and $\phi_\alpha + \theta_{\alpha, i}$ of the primary and the
secondary fibers, as well as on the arc-length along the primary fiber (via
$Y_{\alpha_i}$). In contrast to Eq.~(\ref{eq:bary}), which follows from moving
the primary fiber in a fixed environment, Eqs.~(\ref{eq:bary_network}) and
(\ref{eq:deltaz_network}) are derived from a joint movement of all fibers. For
the following we are only interested in the typical magnitude of
$\delta\!z_\alpha^{\rm rel}$, which may be obtained by averaging over the angles
$\phi_\alpha$ and estimating the typical distance between the center of masses
of the intersecting fibers as $Y_\alpha - Y_{\alpha_i} \sim l_f$. We thus find
that $\delta\!z_\alpha^{\rm rel} \propto \gamma l_f$.

By assuming affine displacements of the fiber centers, we have thus succeeded in
reducing the many-body problem of the movement $\delta\!z_\alpha$ of $N_f$
interacting fibers to the case of a single fiber moving the amount
$\delta\!z_\alpha^{\rm rel}\sim \gamma l_f$ relative to its surrounding. The
modulus can thus be calculated from the knowledge of the energy
$W(\delta\!z^{\rm rel})$ calculated in the previous section. From the definition
of the modulus we find that $G\gamma^2/2 = N_f W/V$, where $V$ is the volume of
the system.

It should be made clear that the assumption of affine displacements of the fiber
centers is different from the usual approach of assigning affine deformations on
the scale of the single polymer segment~\cite{mac95,sto05,shi04}. The latter
would lead to deformations $\delta_{\rm aff} \propto \gamma l_s$, proportional
to the length $l_s$ of the segment.  Instead, axial displacements of the fiber
as a whole are, by construction of the floppy mode, directly translated into
{\it non-affine} deformations $\delta_{\rm na} \propto \gamma l_f$, which do not
depend on the length of the segment but rather on the scale of the fiber length
$l_f$. We would like to emphasize the subtle difference betweeen ``affine
displacements'' of single points (the fiber centers-of-mass), and ``affine
deformations'' of fiber segments of length $l_s$.

\section{Random Network in 2d}\label{sec:random-network-2d}

Having presented the general concepts of the floppy modes and their
energy we now proceed to introduce a simple model system where the
ideas may be tested.  The random two-dimensional network, the ``Mikado
model'' \cite{frey98}, has the advantage that it only needs one
structural parameter, the density of fibers $\rho$.

The network is defined by randomly placing $N$ elastic fibers of
length $l_f$ on a plane of area $A=L^2$ such that both position and
orientation are uniformly distributed~\cite{ast00,ast94,lat01}. The
fiber-fiber intersections are assumed to be perfectly rigid, but
freely rotatable crosslinks that do not allow for relative sliding of
the filaments. The randomness entails a distribution of angles
$\theta\in[0,\pi]$ between two intersecting filaments
\begin{equation}\label{eq:angleDist}
P(\theta) = \frac{\sin\theta}{2}\,,
\end{equation}
while distances between neighbouring intersections, the segment
lengths $l_s$, follow an exponential distribution~\cite{kal60}
\begin{equation}\label{eq:segDist}
  P(l_s)=\lsbar^{-1} e^{-l_s/\lsbar}\,.
\end{equation}
The mean segment length $\lsbar$ is inversely related to the line
density $\rho=Nl_f/A$ as $\lsbar= \pi / 2\rho $. The segments are
modeled as classical beams with cross-section radius $r$ and bending
rigidity $\kappa$. Loaded along their axis (``stretching'') such
slender rods have a rather high stiffness
$k_\parallel(l_s)=4\kappa/l_sr^2$, while they are much softer with
respect to transverse deformations $k_\perp(l_s)=3\kappa/l_s^3$
(``bending'').

Numerical simulations~\cite{frey98,wil03,hea03a} for the effective shear modulus
$G$ of this network have identified a cross-over scaling scenario characterized
by a length scale
\begin{eqnarray}
  \xi=l_f(\delta\!\rho l_f)^{-\nu}
\end{eqnarray}
with $\nu\approx2.84$~\cite{wil03}~\footnote{The density
  $\delta\!\rho$ is measured relative to the rigidity percolation
  transition occuring at $\rho_c l_f=6.7$} that mediates the
transition between two drastically different elastic regimes. For a
fiber radius $r \gg \xi$ the system is in an affine regime where the
elastic response is mainly dominated by stretching deformations
homogeneously distributed throughout the sample. The modulus in this
regime is simply proportional to the typical stretching stiffness,
$G_{\rm aff}\propto k_\parallel(\lsbar)$ and independent of the fiber
length $l_f$.  This is in marked contrast to the second regime at
$r\ll \xi$.  There, only non-affine bending deformations are excited
and the modulus shows a strong dependence on fiber length,
\begin{eqnarray}
G_{\rm na} \propto
k_\perp(\lsbar) \, \left( \frac{l_f}{\lsbar} \right)^{\mu-3} \, , 
\end{eqnarray}
and thus on density, $G_{\rm na}\propto \delta\!\rho^\mu$ where
$\mu=2\nu+1\approx 6.67$.

\begin{figure}[h]
 \begin{center}
   \includegraphics[width=0.9\columnwidth]{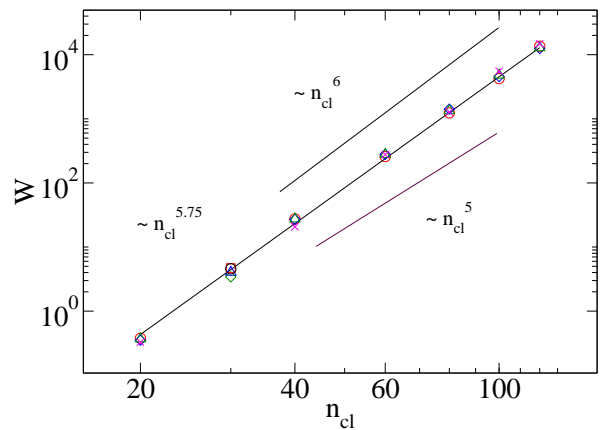}
\end{center}
\caption{Solution of Eqs.~(\ref{eq:selfconsEn}) and (\ref{eq:Wharmonic}) for
  various numbers $n_{\rm cl}$ of crosslinks per filament. The randomness is
  defined by Eqs.(\ref{eq:angleDist}) and (\ref{eq:segDist}). The different
  symbols at given $n_{\rm cl}$ relate to ensembles of varying size $N=100\ldots
  1000$. The lines $W\sim n_{\rm cl}^{5}$ and $W\sim n_{\rm cl}^{6}$ serve to
  illustrate the quality of the fit.}
  \label{fig:k_density}
\end{figure}

As this latter non-affine regime is characterized by a ratio
$k_\parallel(\lsbar)/k_\perp(\lsbar)\sim (\lsbar/r)^2 \gg 1$, and therefore a
bending mode that is soft as compared to the stretching mode, we may apply the
floppy-mode picture developed in previous sections to calculate the exponent
$\mu$. To this end, we numerically solve Eq.~(\ref{eq:selfconsEn}) for varying
numbers $n_{\rm cl}\sim \rho l_f$ of crosslinks per fiber. The average $\langle
.\rangle$ is thereby defined in terms of the probability distributions of
Eqs.~(\ref{eq:angleDist}) and (\ref{eq:segDist}). As a result, we find the fiber
energy
to scale as $W\sim n_{cl}^x\kappa/l_f^3$ and $x\approx5.75$ (see
Fig.~\ref{fig:k_density}). The shear modulus is infered from $W$ as $G = 2\rho
W/l_f\gamma^2\sim \rho^{6.75}$, which reproduces the exponent $\mu$ as measured
in the simulation to a remarkable accuracy.

Based on the formalism of the preceding sections we have also
developed~\cite{heu06floppy} a scaling argument that allows approximate solution
of Eq.~(\ref{eq:selfconsEn}) in terms of a single length-scale $l_{\min}$, which
on a microscopic scale governs the coupling of the fiber to the matrix. Since
the stiffness $k_\perp\sim\kappa/l_s^{3}$ of the individual polymer segment is
strongly increasing with decreasing its length $l_s$, we assume that segments
with $l_s<l_{\rm min}$ rather deform the surrounding medium than being deformed
itself, while longer segments $l_s>l_{\rm min}$ are not stiff enough to deform
the medium. The scale $l_{\rm min}$ therefore plays the role of a minimal length
below which segments are stiff enough to remain undeformed.

In terms of the crosslink deflections $y_i$, this implies that long (and soft)
segments have $y_i \approx \bar y_i$, while short (and stiff) segments have
crosslinks that are in their original position $y_i\approx 0$. Since the energy
of a segment of length $l_s$ can be written as $w(l_s)\sim k_\perp
y_i^2\sim\kappa y_i^2/l_s^3$, we find that the
elastic energy is reduced by the amount $w(l_{\min})\sim\kappa\bar
y_i^2/l_{\min}^3$ as compared to the situation where also the short segments are
deformed. In turn, the energy in the neighbouring fiber is increased, where a
floppy mode of amplitude $\delta\!z\sim \bar y_i$ is excited. The length-scale
$l_{\min}$ can therefore be determined by equating the energy reduction in the
small segments, $w(l_{\min})$, with the energy increase due to the additional
floppy mode in the neighbouring fiber. This latter contribution can be
calculated as an average over all segments of length $l_s>l_{\min}$ thus giving
\begin{equation}\label{eq:avgFiberEnergy}
  W \simeq n_{\rm cl}\int_{l_{\min}}^\infty dl_sP(l_s)w(l_s) \overset{!}{=} w(l_{\rm min})
  \,.
\end{equation}
As a result, we find $l_{\min} \simeq 1/\rho^2l_f$ and thus for the average
fiber energy $W \simeq \kappa(\rho l_f)^6/l_f$. This corresponds to an exponent
$\mu=7$, which confirms the previous analysis.

{F}rom Eq.~(\ref{eq:segDist}) one may also induce a probabilistic
interpretation of the length-scale $l_{\min}$. Segments with lengths
$l_s<l_{\rm min}$ will occur on average only once along a given fiber.
This may be seen from solving the equation
\begin{equation}\label{eq:only_once}
  \int_0^{l_{\rm min}} dl_sP(l_s) \sim \frac{1}{n_{\rm cl}}\,,
\end{equation}
stating that small segments will occur once in every $n_{\rm cl} \sim
\rho l_f$ crosslinks.  There will therefore be typically one segment
per fiber in the undeformed configuration $y_i\approx0$, while all
others follow the floppy mode.

{ These scaling arguments also provide additional insights into a more
  microscopic understanding of the crossover from affine to non-affine
  elasticity in random rod networks \cite{wil03}. Upon rewriting the crossover
  scale $\xi$ as $\xi \sim l_f(l_{\rm min}/l_f)^{3/2}$ the scaling variable
  $x=r/\xi$ of Eq.~(2) in \cite{wil03} takes the alternative form $x^{-2} \sim
  k_\parallel(l_f)/k_\perp(l_{\rm min})$ such that the crossover scaling law of
  the modulus reads
\begin{equation}\label{eq:scalingWilhelm}
  G(r,\rho) = \rho^\mu g(k_\parallel(l_f)/k_\perp(l_{\rm min})) \,.
\end{equation}
The scaling argument now identifies a competition between the relative stiffness
of the stretching and bending modes on the scale of the {\it whole polymer
  fiber} as the driving force of the affine to non-affine crossover. For large
scaling arguments, $k_\parallel(l_f) \gg k_\perp(l_{\rm min})$, bending is the
weaker mode and yields a non-affine response in the form of floppy modes.
Stretching deformations become dominant as soon as $k_\parallel(l_f)$ becomes
smaller than $k_\perp(l_{\rm min})$; this happens if the rigidity scale
$l_\text{min} \leq (r^2 l_f)^{1/3}$.

The only requirement for the presence of a bending dominated regime (beyond the
scale separation $k_\parallel/k_\perp\gg 1$) is a low coordination number, which
for the random fiber network can be calculated as $z=4(1-(\rho l_f)^{-1})$. This
places the network below the rigidity transition for any finite $l_f$, while
increasing the filament length $l_f\to\infty$ the critical coordination of
$z_c=4$ is asymptotically reached. As an implication the bending mode must
eventually be suppressed.

The above analysis clearly shows that the proposed floppy mode concept can be
utilized to understand the bending dominated elasticity in the random fibrous
network. It allows to extract the length-scale $l_{\min}$ that is ultimately
responsible for the strong density-dependence of the elastic modulus as found in
the simulations.  Most importantly, the length-scale $l_{\min}$ is a special
feature of the random architecture studied here. Other network structures will
not necessarily feature the same length-scale even though the basic formalism of
the floppy bending modes can still be applied. The exponents characterizing the
elastic response will thus depend on network architecture, a fact which is also
exemplified in the Appendix.

In Ref.~\cite{lieleg} we have furthermore applied the theory to explain the
mechanics of reconstituted actin networks, where filaments are crosslinked and
bundled by fascin. By taking into account the fact that bundles have to be
characterized by a length-dependent bending rigidity
$\kappa(L)$~\cite{bathe06,heussinger07bundle,claessens06} it was possible to
explain the observed dependence of the elastic modulus on actin and fascin
concentration.


\subsection{Nonlinear elasticity arising from geometric effects}\label{sec:nonlinear}

Here, we report on additional simulations probing the nonlinear modulus of the
structure. Note, that in these simulations the material properties of the fibers
remain linear, such that the nonlinearities result from geometrical effects
only. As one can infer from Fig.~\ref{fig:nonlinear} the network is strongly
stiffening already at very small values of strain.
\begin{figure}[h]
 \begin{center}
   \includegraphics[width=0.9\columnwidth]{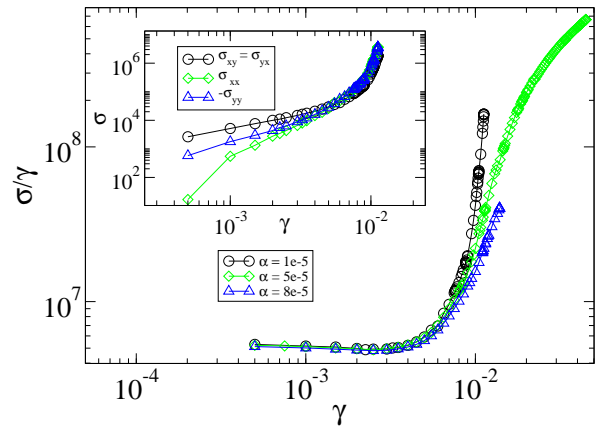}
\end{center}
\caption{Nonlinear ``modulus'' $\sigma/\gamma$ in the bending dominated regime
  ($\rho l_f=30$) for various values of the aspect ratio $\alpha=r/l_f$. Inset:
  The stress increases linearly up to a strain of about $1\%$. Normal stresses
  quickly rise in magnitude and eventually are of the same order and
  proportional to the shear stress.}
  \label{fig:nonlinear}
\end{figure}
Similar results have recently been reported in~\cite{onck05}, where the
stiffening behaviour was attributed to a crossover from bending to stretching
dominated elasticity. The floppy mode picture allows to give this crossover a
microscopic explanation. As argued in Section~\ref{sec:floppy-modes-fibrous},
the floppy modes of the fibrous network are only adequate for infinitesimally
small displacements $\delta\!z$. The construction embodied in Eq.~(\ref{eq:bary})
keeps segment lengths invariant to first order in $\delta\!z$ only, such that
any finite deformation will necessarily lead to stretching of the bonds.

Note, that this stiffening mechanism is not mediated by non-linear material
properties of the fibers but rather is of geometric origin and is due to the
specific structural arrangement in the fibrous architecture. It is therefore of
different nature than the stiffening mechanism inherent to single semiflexible
polymers, where an applied tension can stretch the polymer only as far as there
is stored length available~\cite{marko95}.

In the nonlinear regime we have also measured the normal stresses $\sigma_{xx}$
and $\sigma_{yy}$ that act perpendicular to the principal strain direction. We
found (see inset to Fig.~\ref{fig:nonlinear}) that these stresses can become of
the order of the shear stresses $\sigma_{xy}$ and have a negative sign
indicating that the network ``pulls in'' during the course of the deformation. A
similar effect has recently been observed in rheological measurements on F-actin
networks~\cite{janmey07normalstress} and rationalized in terms of the highly
nonlinear entropic stretching response of single polymers. Note, that in our
simulations the same effect occurs within a purely mechanical picture, where no
material non-linearities are present. It is explained with the fact, that the
additional amount of contour length necessary to undergo a finite floppy mode
can only come from pulling in the fiber ends. This is equivalent to a network
contraction which leads to the observed large normal stresses.

\subsection{Nonstraight fibers}

In real networks fibers will never be perfectly straight. We have argued above
that in this case the scale of the fiber-length $l_f$ must be viewed as the
length-scale over which the polymer remains straight. With this in mind our
theory also holds for networks where fibers are non-straight, as long as the
undulation wavelength $\lambda\sim l_f$ is larger than the distance between
crosslinks $\lsbar$.

In this section we investigate the effects of introducing undulations with
wavelengths comparable to the crosslink distance, $\lambda\sim \lsbar$. To this
end we have manually generated zig-zag fibers by randomly displacing the
crosslinks by some maximal amount $\Delta\cdot l_f$. A similar analysis has
been performed in Ref.~\cite{onck05}, where a substantial decrease in the degree
of non-affinity of the deformation field has been found. Similarly, we find that
the system develops a new crossover to a regime of affine bending deformations
(see Fig.~\ref{fig:CrossOverAffine}), where the modulus scales as
$G\propto\delta\!\rho^3$, a behaviour well known from bending dominated cellular
foams~\cite{gib99,kra94,heu07architecture}.

\begin{figure}[h]
  \begin{center}
    \includegraphics[width=0.9\columnwidth]{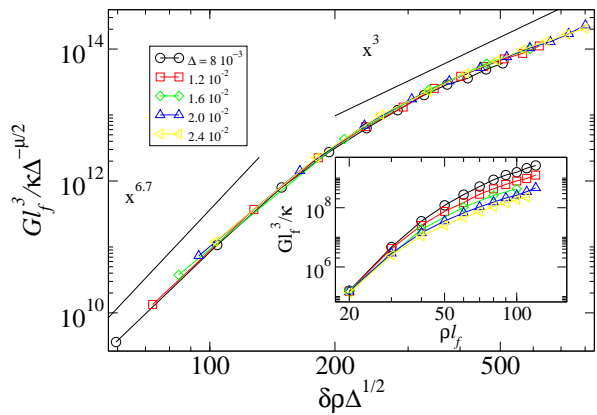}
  \end{center}
  \caption{Shear modulus $G$ (inset) and scaling function $g$ of
    Eq.~(\ref{eq:ModulusCrossOverAffine}) for various values of $\Delta$.
    Collapse is achieved by plotting $G\Delta^{\mu/2}$ as a function
    $x=\delta\!\rho\Delta^{1/2}$ and known exponent $\mu=6.7$~\cite{wil03}. The
    asymptotic regimes show the scaling properties of straight fibers, $g(x)\sim
    x^{6.7}$, and foams, $g(x)\sim x^3$, respectively.}
  \label{fig:CrossOverAffine}
\end{figure}

In this new regime the bending deformations come from pulling out the zig-zags
similar to the pulling of thermally activated polymer undulations. We find that
the curves may be scaled by using the same length-scale $l_{\rm
  min}\sim\delta\!\rho^{-2}$ that served as a lower cut-off in segment lengths. The
modulus thus takes the following scaling form
\begin{equation}\label{eq:ModulusCrossOverAffine}
G(\Delta,\delta\!\rho)=\Delta^{-\mu/2} g(\Delta/l_{\rm min})\,,
\end{equation}
where the scaling function has the limiting form $g(x\ll 1)\sim x^{\mu/2}$ to
eliminate the $\Delta$-dependency. For large values of the scaling variable
$x\gg1$ we have to recover the scaling properties characteristic for foams,
giving $g(x \gg 1)\sim x^{3/2}$. This analysis highlights once more the
fundamental role played by the length-scale $l_{\min}$ in establishing the
elastic response of the network. Here, it acts as a crossover scale, that
mediates the transition to a foam-like bending regime at strong disorder $\Delta
\gg l_{\min}$.

Note, that by introducing kinks in the contour of the fibers, the floppy modes
start to spread beyond the single fiber to which they were confined originally.
A kink is most conveniently characterized by the angle $\psi$ through which the
direction of the fiber changes at the location of the kink. By displacing a
crosslink by the amount $\Delta$ one thus finds for the angle
$\sin\psi=\Delta/l_s$, where $l_s$ is the length of the segment that ends at
the crosslink. Exciting the fiber with a floppy mode of amplitude $\delta\!z$,
a finite kink-angle $\psi$ leads to the fraction $\delta\!z'\sim
\delta\!z\sin\psi \sim \delta\!z\Delta/l_s$ being coupled into the
neighbouring fiber. At the crossover, defined by $x=\Delta/l_{\min}\sim 1$, we
therefore find that for a segment of length $l_s=l_{\min}$, $\delta z'(l_{\min})
\approx \delta z$.  In this situation the floppy mode is transmitted to the
neighbouring fiber without attenuation of its amplitude. Since segments of
length $l_{\min}$ statistically occur once per filament, the crossover point
also marks the onset of a complete delocalization of the floppy modes.

\section{Conclusion}

We started our discussion with the assumption that the elasticity of stiff
polymer networks is governed by the action of the bending mode. This assumption
is based on the recognition that in systems where the persistence length is
large, bending as compared to stretching is by far the softer mode. The
respective spring constants are scale-separated and obey the relation
$k_\parallel/k_\perp \sim l_p/l \gg 1$.

One immediate implication of this scenario is that polymer end-to-end distances
have to stay constant, which necessitates deformations that are highly
non-affine. We have characterized this non-affine deformation field by
constructing the floppy modes of the structure~\cite{heu06floppy}. These are
defined as the set of crosslink displacements that do not lead to any
stretching of bonds. With this microscopic deformation field it is possible to
calculate the macroscopic elastic moduli on the level of a self-consistent
effective medium theory that incorporates fiber-medium interactions within a
Cayley-tree approximation.

As a result the anomalous scaling properties of the linear shear modulus as
determined by computer simulations of two-dimensional random
networks~\cite{wil03,hea03a} are explained. The exponents are found to be a
consequence of the special architecture of the network that features two
different length-scales. On the mesoscopic scale the fiber length $l_f$ induces
a non-affine deformation field, with segment deformations $\delta_{\rm na}$
following the macroscopic strain $\gamma$ as $\delta_{\rm na}\sim \gamma l_f$,
instead of as $\delta_{\rm aff}\sim\gamma l_s$, which would result from an
affine deformation field.  Microscopically, a second length $l_{\min}$ plays the
role of a minimal length below which segments are stiff enough to remain
undeformed.

We would like to emphasize that the construction of the floppy modes only relies
on the presence of the mesoscopic length $l_f$, which is applicable to a broad
class of networks. In the particular case of a random rod network we have found
that the anomalous scaling properties of the shear modulus, previously found in
simulations, crucially depend on the presence of a second length-scale
$l_{\min}$, which is a special property of this random architecture.  The
exponents found for random rod networks are therefore not immediately applicable
to other systems. Having established the general theoretical framework, it is
nevertheless straightforward to calculate the exponents for other types of
networks in two and three dimensions.  Indeed, we have applied the theory to
reconstituted actin networks crosslinked and bundled with fascin, and found
that the calculated exponents are in good agreement with the experimental
results~\cite{lieleg}.

Finally, we also conducted simulations probing the non-linear elasticity of the
random fibrous network as well as modified the network structure by introducing
kinks in the contour of the polymers. The results confirm the governing role of
the identified length-scales and firmly establish that the non-affine floppy
mode picture captures the essential physics of stiff polymer networks similar in
spirit to affine rubber elasticity for flexible polymer gels. In view of this
conceptual analogy, the next step could be to assess the importance of
crosslink fluctuations, which have been neglected here (as in classical rubber
elasticity). By greatly reducing the number of fluctuating degrees of freedom to
one per fiber (namely $\delta\!z$), the theory developed here may very well
provide a new starting point for the analysis of the statistical mechanics of
stiff polymer networks.

\begin{acknowledgments}
  We gratefully acknowledge fruitful discussions with Mark Bathe and Camilla
  Mohrdieck. Financial support from the German Science Foundation (SFB 486) and
  the German Excellence Initiative via the program "Nanosystems Initiative
  Munich (NIM)" is gratefully acknowledged.
\end{acknowledgments}

\appendix

\section{Solution of Eq.~(\ref{eq:selfconsEn}) for various network structures}

In this appendix we provide some technical details on how to solve
Eq.~(\ref{eq:selfconsEn}) for various network architectures in two spatial
dimensions. We will measure lengths in units of the fiber length $l_f$ and
energies in units of $\kappa/l_f$, where $\kappa$ is the bending stiffness of
the fiber. Assuming harmonic energies $W(x) = kx^2/2$ we rewrite
Eq.~(\ref{eq:selfconsEn}) symbolically as
\begin{equation}\label{eq:selfconsSymbol}
  k=\langle f(k,n_{\rm cl};\{z_i,\theta_i\}) \rangle\,,
\end{equation}
where the function $f$ is defined by
\begin{equation}\label{eq:selfconsk}
  f = \min_{y(z)}\left( \frac{2W_b}{\delta\!z^2} +
    k\sum_{i=1}^{n_{\rm cl}}
    \sin^2(\theta_i)\left(\frac{y(z_i)}{\delta\!z}+\cot\theta_i\right)^2 \right)\,,
\end{equation}
and we used Eq.~(\ref{eq:bary}) to substitute $\bar
y_i=-\cot\theta_i\delta\!z$.

The network structure enters Eqs.(\ref{eq:selfconsSymbol}) and
(\ref{eq:selfconsk}) via the variables $\{z_i,\theta_i\}$, which relate to the
locations $z_i$ of the crosss-links on the backbone of the primary fiber as well
as the angles $\theta_i$ between primary and secondary fibers. The ensemble
average $\langle.\rangle$ can then be defined by the probability distributions
$P(\{\theta_i\})$ and $P(\{l_i\})$, where segment lengths are given by $l_i =
z_{i+1}-z_i$.

To illustrate the importance of structural features on the elastic
properties of the network we solve Eq.~(\ref{eq:selfconsk}) for two types of
distributions, relating to random and regular structures, respectively.  The
random network is characterized by probability distributions as given in
Eqs.~(\ref{eq:angleDist}) and (\ref{eq:segDist}). The regular network has only
one segment length $l_0=l_f/(n_{\rm cl}-1)$ and an angular distribution similar
to Eq.~(\ref{eq:angleDist}) but restricted to the interval
$[\theta_{\min},\pi-\theta_{\min}]$.

For a given realization of the randomness the function $f$ is calculated by
performing the minimization with respect to the contour $y(z)$. This is achieved
in two steps, where first the bending energy $W_b[y]$ is minimized for a
\emph{given} set of values $\{y(z_i)\}$. As explained in the main text, this is
equivalent to a cubic spline interpolation. The second step consists of a
minimization with respect to the remaining variables $\{y(z_i)\}$.

Finally solving Eq.~(\ref{eq:selfconsSymbol}) the fiber stiffness
$k^\star(n_{\rm cl})$ is determined as a function of the number $n_{\rm cl}$ of
crosslinks per fiber. A graphical solution for $n_{\rm cl}=40$ for various
network structures is presented in Fig.~\ref{fig:vardistr}. The function
$\langle f(k)\rangle$ is plotted as function of $k$. The sought after value
$k^\star$ is found at the point of intersection with the bisecting curve.

\begin{figure}[h]
 \begin{center}
  \includegraphics[width=0.9\columnwidth]{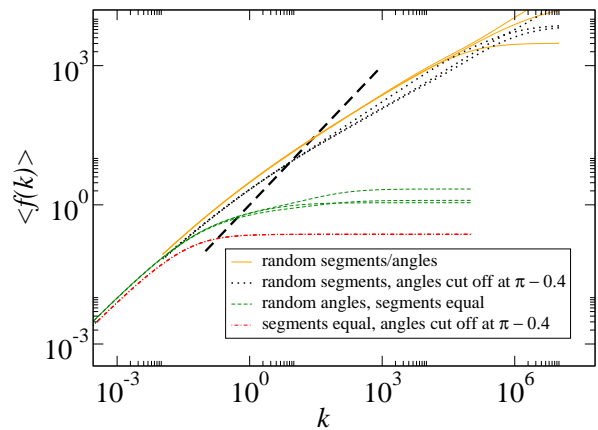}
\end{center}
\caption{Graphical solution of Eq.~(\ref{eq:selfconsk}) for a fiber with
  $n_{cl}=40$ crosslinks imbedded in networks of varying architectures. $\langle
  f(k)\rangle$ is plotted as a function of $k$ for ensembles of varying sizes.
  The solution to Eq.~(\ref{eq:selfconsk}) is found by intersecting the curves
  with the bisecting line (dashed).}
  \label{fig:vardistr}
\end{figure}

The different curves for a given network structure correspond to
ensembles of varying size.  They seem to diverge in the limit $k\to\infty$. In
fact, in this limit only the bending energy $W_b$ contributes to
Eq.~(\ref{eq:selfconsk}) and $y_i\approx \bar y_i$.  This may make the averaging
procedure ill defined, for example in the case of Eq.~(\ref{eq:segDist}) where
the segment lengths $l_s$ can become arbitrarily small. The resulting segmental
bending energy $w_b\sim l_s^{-3}$ shows a divergence and does not have a well
defined average value.

As one can see from Fig.~\ref{fig:vardistr} the resulting fiber stiffness
$k^\star$ very sensitively depends on the randomness in the segment lengths,
while crosslink angles only play a minor role. This is made particularly clear
by comparing the random-segment and the regular-segment network in terms of the
exponent $x$, which is defined by $k^\star \sim n_{\rm cl}^x$. As stated in the
main text the random network has $x\approx 6$, while for the regular network we
find $x\approx4$. We have shown above that the former result derives from the
presence of the length-scale $l_{\rm min}$. In contrast, the latter is simply
obtained by calculating the bending energy of $n_{\rm cl}$ segments each of
length $l_0$, $W_b\sim n_{\rm cl}\kappa/l_0^{3}\sim n_{\rm cl}^4$.


\end{document}